\documentclass[nofootinbib,aps,amssymb,floatfix,superscriptaddress,preprintnumbers]{revtex4}

\usepackage{graphicx}
\usepackage{bm}
\usepackage{amsmath}
\usepackage{amssymb}
\usepackage{amsfonts}
\usepackage{float}
\usepackage{hyperref}
\usepackage{dsfont}  
\usepackage{slashed}  
\usepackage{booktabs}
\usepackage{multirow}
\usepackage{subfigure}
\usepackage[sort&compress]{natbib}
\usepackage{xcolor}
\usepackage{ulem}

\newcommand{\be}{\begin{equation}}  
\newcommand{\ee}{\end{equation}}  
\newcommand{\beq}{\begin{eqnarray}} 
\newcommand{\eeq}{\end{eqnarray}}

\newcommand{\bea}{\begin{eqnarray}}
\newcommand{\eea}{\end{eqnarray}}

\newcommand{\Slash}[1]{{\ooalign{\hfil/\hfil\crcr$#1$}}}
\newcommand{\nn}{\nonumber \\}

\begin{document}
\preprint{ZTF-EP-25-05}

\title{Directed flow from parton spin-orbit coupling in $pp$ and $pA$ collisions }

\author{Sanjin Beni\'{c}}
\affiliation{Department of Physics, Faculty of Science, University of Zagreb, Bijenička c. 32, 10000 Zagreb, Croatia}

\author{Yoshitaka Hatta  }
\affiliation{Physics Department, Brookhaven National Laboratory, Upton, NY 11973, USA}
\affiliation{RIKEN BNL Research Center, Brookhaven National Laboratory, Upton, NY 11973, USA}

\begin{abstract}

We point out a novel mechanism to generate  $\cos \phi$ two-particle azimuthal correlation (`directed flow') in unpolarized proton-proton and proton-nucleus collisions  in the forward rapidity region of the projectile proton. This is a direct consequence of the recently discovered strong spin-orbit coupling in gluons at small-$x$. The observable simultaneously serves as a unique probe into the double helicity parton distribution functions of the proton.

\end{abstract}

\maketitle

\section{Introduction}

Spin-orbit coupling $\vec{L}\cdot \vec{S}$ between spin $\vec{S}$ and orbital angular momentum (OAM)  $\vec{L}$  is a ubiquitous phenomenon across many areas of contemporary sciences such as condensed matter physics, atomic physics and chemistry. In nuclear physics, it is well established that the strong spin-orbit coupling acting on nucleons inside  a nucleus is the key to explaining the nuclear magic numbers  \cite{mayer}. Moreover, it is possible to directly probe the strength of the nucleon spin-orbit coupling in low-energy scattering experiments as described  in a classic paper by Fermi \cite{fermi}.  Once a proton moving in the $z$-direction and polarized in the $y$-direction enters the interior of a target nucleus, it feels the spin-orbit potential   $V\sim \vec{L}\cdot \vec{S}$ of the nucleus. The initial proton spin $S_y$ is then converted to a kick in the $x$-direction $L_y\sim -xp_z$, generating asymmetry between  $\pm x$ directions. Experimentally, this can be observed as  analyzing power $A_N$, also known as  transverse single spin asymmetry. 
 Historically, such experimental data have been used to determine  the spin-orbit potential as a part of the nuclear optical potential \cite{Varner:1991zz}.

Quarks and gluons orbiting inside a nucleon also feel spin-orbit coupling, but its theoretical description is more subtle due to the complexity of QCD as a non-Abelian gauge field theory. A consistent formulation is possible in the infinite momentum frame where one can describe quarks and gluons as quasifree particles (`partons') carrying a longitudinal momentum fraction $x$ of the parent nucleon and helicity $S_z^{q,g}$. One can then define the `spin-orbit correlation' $C_{q,g}(x) \sim \langle L^{q,g}_zS^{q,g}_z\rangle$ \cite{Lorce:2011kd,Kanazawa:2014nha,Mukherjee:2014nya,Engelhardt:2021kdo,Bhattacharya:2024sno,Hatta:2024otc,Bhattacharya:2024wtg} as an analog of $\vec{L}\cdot \vec{S}$ in nonrelativistic systems. Recent studies   \cite{Bhattacharya:2024sno,Hatta:2024otc} have  shown that, in the small-$x$ region $x\ll 1$,  $C_{q,g}(x)$ are simply related to the unpolarized parton distribution functions (PDFs) $C_q(x)\approx -\frac{1}{2}q(x)$ and $C_g(x)\approx -g(x)$. Physically, this means that the helicity and OAM of individual quarks and gluons are perfectly  anti-aligned. Such states can be interpreted as Bell states representing a maximal entanglement between helicity and OAM   in the quantum mechanical sense   \cite{Bhattacharya:2024sno,Hatta:2024lbw,Agrawal:2025yoe}. 

It is then very interesting to directly probe $C_{q,g}(x)$ in high energy experiments. However, this turns out to be nontrivial. Naively, in order to probe  the operator  $L_zS_z$, the incident proton needs to be longitudinally polarized. The quark (or gluon) helicity $S_z$ inside the polarized proton is then converted to $L_z=xp_y-yp_x$. Since  $\langle p_{x,y}\rangle =0$,  asymmetries in the $xy$ plane cannot be generated. This is related to  the fact that longitudinal single spin asymmetry  in single particle production vanishes in QCD. Nevertheless, observables sensitive to $C_{q,g}(x)$ have been identified  in the literature by looking at final states with at least two identified particles (jets) \cite{Bhattacharya:2017bvs,Boussarie:2018zwg,Echevarria:2022ztg,Bhattacharya:2023hbq,Bhattacharya:2024sck}. However, in these instances, there are background contributions of comparable  size which make the extraction of $C_{q,g}(x)$ complicated. 

The main idea of this paper is that one can access the gluon spin-orbit correlation $C_g(x)$ at small-$x$ in {\it double} parton scattering. Indeed, if there are two quarks in the initial state, $S^1_zS^2_z$ is converted to $L_z^1L_z^2=(\vec{r}_1\times \vec{p}_1)_z(\vec{r}_2\times \vec{p}_2)_z \sim (\vec{r}_1\cdot \vec{r}_2) (\vec{p}_1\cdot \vec{p}_2)$, and this can be nonvanishing. The experimental signal is then  $\langle \vec{p}_1\cdot \vec{p}_2\rangle \sim \cos (\phi_1-\phi_2)$ two-particle azimuthal angular correlation. In the context of heavy-ion collisions, such a  correlation is called `directed flow' $v_1$ that can be attributed to the  hydrodynamic evolution of the quark gluon plasma  \cite{Teaney:2010vd}. Interestingly, $v_1$ has also been observed in high-multiplicity proton-nucleus ($pA$) and deuterion-nucleus ($dA$) collisions around midrapidity at the LHC  \cite{ATLAS:2014qaj} and at RHIC \cite{STAR:2015kak}, although its interpretation is not as straightforward. In the present work, we shall focus on $v_1$ in the forward rapidity region of  proton-proton ($pp$) and $pA$ collisions which should be unambiguously interpreted as cold QCD effects. In this context, previously $v_1$ was shown to arise from the odderon exchange in QCD \cite{Boer:2018vdi} (see also \cite{Soudi:2024slz}). We will contrast our finding with this reference. 

As is already implied in the above argument, our observable simultaneously probes the double {\it helicity} parton distribution function (PDF) $F_{\Delta q\Delta q}\sim \langle S^q_zS^q_z\rangle \sim \langle (\bar{q}\gamma^+\gamma_5q)^2 \rangle$ of the projectile proton which is nonvanishing even if the proton is unpolarized.  In contrast to the double unpolarized PDF $F_{qq}\sim  \langle (\bar{q}\gamma^+ q)^2 \rangle$ \cite{Gaunt:2009re,Diehl:2011yj},  the discussion of $F_{\Delta q\Delta q}$  has been quite limited \cite{Diehl:2011yj,Chang:2012nw,Rinaldi:2014ddl},  especially in connection with phenomenology. Typically, in processes where  $F_{qq}$ is  relevant, $F_{\Delta q\Delta q}$  also appears as a subleading correction \cite{Kasemets:2012pr,Echevarria:2015ufa}.  
As we shall see, the present observable should allow  direct access to   $F_{\Delta q\Delta q}$ in the valence region after subtracting backgrounds.

\section{Double parton scattering}

Consider high energy unpolarized proton-proton ($pp$) or proton-nucleus ($pA$) collisions. The projectile proton  moves in the $+z$ direction with momentum $P^\mu \approx \delta^\mu_+P^+$ and the target proton/nucleus moves in the $-z$ direction with momentum $P'^\mu \approx \delta^\mu_-P'^-$. The center of mass energy is denoted by $s\approx 2P^+P'^-$. We shall be interested in particle production in the forward rapidity  region $y>0$ of the projectile. When $y\gg 1$, the process probes  the small-$x$ region of the target densely populated by gluons. In the hybrid factorization approach \cite{Dumitru:2005gt} appropriate for this kinematics, the squared scattering amplitude  between a collinear quark $p^\mu=(xP^+,0,{\bm 0})$ from the projectile and a gluon from the target with transverse momentum $p'^\mu=(0,x'P'^-,{\bm k})$  reads
\beq
(\bar{q}\gamma_j\Slash p_f\gamma_iq) F^{-j}F^{-i} &\approx & \frac{1}{4}\left({\rm Tr}[\gamma_j\Slash p_f\gamma_i \gamma^-]\bar{q}\gamma^+q +   {\rm Tr}[\gamma_j\Slash p_f\gamma_i\gamma_5\gamma^-]\bar{q}\gamma^+\gamma_5q\right) F^{-j}F^{-i} \nn 
&\propto& \bar{q}\gamma^+q \,  F^{-\mu}F_\mu^{\ -}+ \bar{q}\gamma^+\gamma_5q 
\, iF^{-\mu}\tilde{F}_\mu^{\ -}, \label{first}
\eeq
where $p_f^\mu=(xP^+,x'P'^-,{\bm k})$ is the final state quark momentum which is on-shell. The first term represents  
the product of the unpolarized quark PDF in the projectile and the unpolarized gluon transverse momentum dependent distribution (TMD) in the target. The second term features   the  polarized quark PDF in the projectile. The latter is irrelevant in unpolarized scattering for single-particle production.  The differential cross section is given by 
\beq
\frac{d\sigma}{dy d^2{\bm P}} =\int_{z_{\rm min}} \frac{dz}{z^2} D(z) xq(x) \frac{2\pi^2\alpha_s}{N_c}\frac{F(x',{\bm k})}{{\bm k}^2}, 
\label{single}
\eeq
where $D(z)$ is the fragmentation function (FF) and ${\bm P}=z{\bm k}$. $F$ is the gluon TMD related to the gluon PDF as $\int d^2 {\bm k} F(x,{\bm k})=xg(x)$. The kinematical variables are related as 
\beq
x'= \frac{{\bm P}^2}{xz^2s}, \qquad  y=\frac{1}{2}\ln \frac{xP^+}{x'P'^-},
\eeq
which sets $z_{\rm min}=|{\bm P}|e^y/\sqrt{s}$ in the center-of-mass frame $P^+=P'^-$. 
If $y$ is sufficiently large for gluon saturation effects in the target to become significant, they can be incorporated by substituting 
\beq
 \frac{2\pi^2\alpha_s}{N_c}\frac{F(x,{\bm k})}{{\bm k}^2} = \int d^2{\bm b}\int \frac{d^2{\bm r}}{(2\pi)^2} e^{-i{\bm k}\cdot {\bm r}}S(x,{\bm r},{\bm b}),
\label{eq:FS}
\eeq
where the dipole $S$-matrix is defined by 
\beq
 2P'^-2\pi\delta(P'^--P''^-) S(x',{\bm r},{\bm b}) = \int \frac{d^2{\bm \Delta}}{(2\pi)^2} e^{-i{\bm \Delta}\cdot {\bm b}} \langle P''|\frac{1}{N_c}{\rm Tr}[U({\bm r}/2)U^\dagger(-{\bm r}/2) ]|P'\rangle,
\eeq
with ${\bm \Delta}={\bm P}''-{\bm P}'$ and $U({\bm r})=P\exp{\left(ig\int_{-\infty}^\infty dz^+A^-(z^+,{\bm r})\right)}$ being the light-like Wilson line. One then recovers  the standard formula  \cite{Dumitru:2005gt}  for single particle production in the Color Glass Condensate (CGC) framework.

Now consider two-particle production in the regime $y_{1,2}\gg 1$. We require $|y_1-y_2|\gtrsim 2$ in order to eliminate  the possibility that the two particles come from the fragmentation of a single parton  when they are close in azimuthal angle $\Delta \phi = \phi_1-\phi_2\approx 0$.   A general picture is that the leading twist $2\to 2$ scattering contribution is peaked around the back-to-back region $|\Delta \phi|\approx \pi$, while double parton scattering gives an approximately $\Delta \phi$-independent (`pedestal') contribution  \cite{Strikman:2010bg,Stasto:2011ru,Lappi:2012nh}. We shall focus on the latter which can be obtained by squaring the first term of  (\ref{first}) \cite{Stasto:2011ru,Hagiwara:2017ofm} 
\beq
\frac{d\sigma_{qq}}{dy_1dy_2 d^2{\bm P}_1d^2{\bm P}_2} &=&\int \frac{dz_1dz_2}{z_1^2z_2^2}D(z_1)D(z_2)\int \frac{d^2{\bm r}_1 d^2{\bm r}_2}{(2\pi)^4}d^2{\bm b}_1d^2{\bm b}_2  e^{-i{\bm k}_1\cdot {\bm r}_1-i{\bm k}_2\cdot {\bm r}_2} \nn 
&& \quad \times x_1x_2F_{qq}(x_1,x_2,|{\bm b}_1-{\bm b}_2|)  S(x'_1,{\bm r}_1,{\bm b}_1) S(x'_2, {\bm r}_2,{\bm b}_2) \nn 
&=& \int \frac{dz_1dz_2}{z_1^2z_2^2}D(z_1)D(z_2)\int \frac{d^2{\bm \Delta}}{(2\pi)^6} x_1x_2F_{qq}(x_1,x_2,|{\bm \Delta}|)  S(x'_1,{\bm k}_1,-{\bm \Delta}) S(x'_2, {\bm k}_2,{\bm \Delta}).
\label{double}
\eeq
A factor of $\frac{1}{2}$ is needed if the two partons are identical. 
$F_{qq}$ is the double quark distribution function \cite{Diehl:2011yj}
\beq
F_{qq}(x_1, x_2, |{\bm b}|) & =& 2 P^+ \int db^- \int \frac{d z_1^-}{2\pi}\frac{d z_2^-}{2\pi} e^{i(x_1 z_1^- + x_2 z_2^-)P^+} \nn
&&\times\langle P|\bar{q}(b - z_1/2)\frac{1}{2}\gamma^+ q(b + z_1/2)\bar{q}(- z_2/2)\frac{1}{2}\gamma^+ q(z_2/2)|P\rangle\,, \label{dp}
\eeq
 where $b=(b^-,{\bm b})$ and $z_{1,2}=(z_{1,2}^-,{\bm 0})$. ${\bm b}$ is the impact parameter between the two quarks. 
In the last line of (\ref{double}) we have introduced the Fourier transforms 
\beq
 S(x',{\bm k},{\bm \Delta})= \int d^2{\bm r}d^2{\bm b} \, e^{i{\bm \Delta}\cdot {\bm b}-i{\bm k}\cdot {\bm r}} S(x',{\bm r},{\bm b}),\qquad F_{qq}(x_1,x_2,|{\bm b}|)= \int \frac{d^2{\bm \Delta}}{(2\pi)^2} e^{-i{\bm \Delta}\cdot {\bm b}} F_{qq}(x_1,x_2,|{\bm \Delta}|).
\eeq 
 In deriving  \eqref{double}, we approximated the  double gluon distribution by the product of single gluon distributions assuming that the summation over intermediate states is saturated by the proton/nucleus  single particle state
\beq
&& 2P'^-\! \int d b^+\int \frac{d^3 
z_1}{(2\pi)^3} \frac{d^3 z_2}{(2\pi)^3}  e^{i (x'_1 z_1^+ + x'_2 z_2^+ )P'^-}e^{ - i {\bm k}_2 \cdot {\bm z}_2 - i {\bm k}_2 \cdot {\bm z}_2}\langle P'| F^{-\mu}(-z_1/2) F^{\ -}_\mu(z_1/2) F^{-\nu}(b-z_2/2)F^{\ -}_\nu(b+z_2/2)|P'\rangle \nn
&& \approx\int \frac{d^2 {\bm \Delta}}{(2\pi)^2} e^{-i{\bm \Delta}\cdot{\bm b}}\int \frac{d^3 
z_1}{(2\pi)^3} e^{i x'_1 P'^- z_1^+ - i {\bm k}_1 \cdot {\bm z}_1}\langle P'| F^{-\mu}(-z_1/2) F^{\ -}_{\mu}(z_1/2) |P' + {\bm \Delta}\rangle\nn
&& \qquad \qquad \times\int \frac{d^3 z_2}{(2\pi)^3} e^{i x'_2 P'^- z_2^+  - i {\bm k}_2 \cdot {\bm z}_2 }\langle P' + {\bm \Delta}|F^{-\nu}(-z_2/2)F^{\ -}_\nu(z_2/2)|P'\rangle \nn
&& = x_1' P'^- x_2' P'^-\int \frac{d^2 {\bm \Delta}}{(2\pi)^2} e^{-i{\bm \Delta}\cdot{\bm b}}f_g(x'_1,{\bm k}_1,-{\bm \Delta})f_g(x'_2,{\bm k}_2,{\bm \Delta})\,,\label{dgluon}
\eeq
where in the last line we recovered the unpolarized gluon generalized TMD (GTMD) $f_g(x,{\bm k},{\bm \Delta})$. At small-$x$, this is related to the off-forward dipole $S(x,{\bm k},{\bm \Delta})$ as 
$x f_g(x,{\bm k},{\bm \Delta}) \approx N_c{\bm k}^2 S(x,{\bm k},{\bm \Delta})/(8\pi^4 \alpha_s)$ \cite{Hatta:2016dxp}, leading to \eqref{double}. In the forward limit $x f_g(x,{\bm k},0) = F(x,{\bm k})$, see \eqref{eq:FS}. In the small-$x$ literature it is common to encounter this result in the context of the large-$N_c$ approximation: $\langle \frac{1}{N_c^2} {\rm Tr}(U U^\dag) {\rm Tr}(U U^\dag)\rangle \approx \langle \frac{1}{N_c}{\rm Tr}(U U^\dag)\rangle\langle \frac{1}{N_c}{\rm Tr}(U U^\dag)\rangle = S S$. As the present derivation shows, this approximation is equivalent to saturating the intermediate states by the single proton/nucleus state. 

 It should be mentioned that, in (\ref{double}), we have neglected a possible contribution from the `color-octet' double parton distribution. This arises from  nontrivial color reconnections in double parton scattering \cite{Mekhfi:1985dv,Diehl:2011yj}. On the projectile side   (suppressing gamma matrix structures),
\beq
\bar{q}_i(t^at^b)_{ij}q_j \, \bar{q}_k(t^ct^d)_{kl}q_l 
= \frac{1}{N_c^2}\bar{q}q \, \bar{q}q{\rm Tr}[t^at^b]{\rm Tr}[t^ct^d] + \frac{4}{N_c^2-1}\bar{q}t^eq\, \bar{q}t^eq {\rm Tr}[t^ft^at^b]{\rm Tr}[t^ft^ct^d]. \nn 
\sim \frac{F_{qq}}{N_c^2}{\rm Tr}[t^at^b]{\rm Tr}[t^ct^d] +
\frac{F_{qq}^8}{N_c\sqrt{N_c^2-1}}\left({\rm Tr}[t^at^bt^ct^d]-\frac{1}{N_c}{\rm Tr}[t^at^b]{\rm Tr}[t^ct^d]\right).
\eeq
Only the first term $F_{qq}\sim \langle \bar{q}q\, \bar{q}q\rangle$ is kept in (\ref{double}). The second term features  the color octet double parton distribution $F_{qq}^8\sim \frac{2N_c}{\sqrt{N_c^2-1}}\langle \bar{q}t^eq\, \bar{q}t^eq\rangle$ that couples to the color quadrupole of the target $\langle{\rm Tr}[UU^\dagger UU^\dagger]\rangle$ in the saturated regime. Since $F_{qq}\gtrsim |F_{qq}^8|$ \cite{Kasemets:2014yna}, the latter is suppressed in the large-$N_c$ counting. We therefore neglect the octet contribution, although it is interesting to investigate its numerical impact  in future work.   

Despite the resulting seemingly factorized structure, (\ref{double}) can exhibit nontrivial azimuthal angular correlations between ${\bm P}_1$ and ${\bm P}_2$.\footnote{It should be mentioned that  correlations  neglected in the  approximation $\langle SS\rangle \approx SS$ in (\ref{double}) can give rise to nontrivial $\Delta\phi$-dependencies  \cite{Hatta:2007fg,Dumitru:2008wn,Kovner:2010xk,Kovchegov:2012nd} including a $\cos \Delta \phi$ component \cite{Dumitru:2008wn}.}  This is because the correlations  between ${\bm r}_{1}$ and ${\bm b}_{1}$ and ${\bm r}_2$ and ${\bm b}_2$ can communicate with each other thanks to the dependence of $F_{qq}$ on $|{\bm b}_1-{\bm b}_2|$. For example, the elliptic gluon distribution $S({\bm r},{\bm b}) \sim \cos 2(\phi_r-\phi_b)$ \cite{Hatta:2016dxp} leads to the $\cos 2(\phi_{P_1}-\phi_{P_2})$ correlation \cite{Levin:2011fb,Hagiwara:2017ofm,Iancu:2017fzn}. More relevant to the present work is the  $\cos (\phi_{P_1}-\phi_{P_2})$ correlation  \cite{Boer:2018vdi}
\beq
d\sigma \sim \cos (\phi_{P_1}-\phi_{P_2}) F_{qq}\otimes O({\bm k}_1)\otimes O({\bm k}_2) ,\label{odd}
 \eeq 
 originating from the  imaginary part of the $S$-matrix ${\rm Im}S({\bm r},{\bm b}) \sim \cos (\phi_r-\phi_b)O({\bm r},{\bm b})$ known as the odderon.\footnote{Recently, Ref.~\cite{Soudi:2024slz} proposed  another  mechanism to generate $v_1$ from certain TMDs. However, numerical results were not presented.}

The central observation of this work is that there exists another, potentially more important source of  $\cos (\phi_{P_1}-\phi_{P_2})$ correlation from the square of the second term in (\ref{first}).  Proceeding as before, we find the following contribution to the cross section
\be
 \frac{d\sigma_{\Delta q \Delta q}}{dy_1dy_2d^2 {\bm P}_{1} d^2 {\bm P}_{2}} = \frac{ 4\pi^4 \alpha_s^2}{s^2 N_c^2} \int \frac{d z_1}{z_1^2} D(z_1) \int\frac{d z_2}{z_2^2} D(z_2) \int \frac{d^2 {\bm \Delta}}{(2\pi)^2} F_{\Delta q \Delta q}(x_1,x_2,|{\bm \Delta|}) \tilde{f}_g(x_1',{\bm k}_1,-{\bm \Delta}) \tilde{f}_g(x_2',{\bm k}_{2},{\bm \Delta})\,.
\label{spinorbxsec}
\ee
$F_{\Delta q \Delta q}$ is the quark double {\it helicity} PDF (in momentum space)  \cite{Diehl:2011yj}
\beq
F_{\Delta q\Delta q}(x_1, x_2, |{\bm \Delta}|) & =&  2 P^+ \int d^3b e^{-i{\bm \Delta}\cdot {\bm b}} \int \frac{d z_1^-}{2\pi}\frac{d z_2^-}{2\pi} e^{i(x_1 z_1^- + x_2 z_2^-)P^+} \nn
&&\times\langle P|\bar{q}(b - z_1/2)\frac{1}{2}\gamma^+\gamma_5 q(b + z_1/2)\bar{q}(- z_2/2)\frac{1}{2}\gamma^+ \gamma_5q(z_2/2)|P\rangle\,,
\label{double2}
\eeq
which is nonvanishing even if the projectile proton is unpolarized. 
$\tilde{f}_g$ is the helicity gluon GTMD  
\beq
\tilde{f}_g(x',{\bm k},{\bm \Delta})=\frac{1}{x' P'^-}\int \frac{d^3z}{(2\pi)^3}e^{ix'P'^-z^+-i{\bm k}\cdot {\bm z}}\langle P'+{\bm \Delta}|iF^{-\mu}(-z/2)\tilde{F}_\mu^{\  -}(z/2)|P'\rangle,
\eeq
which is again nonvanishing for unpolarized targets and features the spin-orbit correlation 
\be
 \tilde{f}_g (x',{\bm k},{\bm \Delta}) 
\approx
-i \frac{{\bm k} \times {\bm \Delta}}{M^2} C_g(x',{\bm k},{\bm \Delta})+\cdots  \,.
\ee
Substituting this into (\ref{spinorbxsec}), we find 
\be
\begin{split}
 \frac{d\sigma_{\Delta q \Delta q}}{dy_1dy_2 d{\bm P}_{1} d^2 {\bm P}_{2}} & = \frac{ 2\pi^4 \alpha_s^2}{s^2 N_c^2} \int \frac{d z_1}{z_1^2} D(z_1) \int\frac{d z_2}{z_2^2} D(z_2) {\bm k}_{1}\cdot {\bm k}_{2} \\
& \times \int \frac{ d^2 {\bm \Delta}}{(2\pi)^2} {\bm \Delta}^{2} F_{\Delta q \Delta q}(x_1,x_2,|{\bm \Delta}|) \frac{C_g(x_1',{\bm k}_{1},-{\bm \Delta})}{M^2} \frac{C_g(x_2',{\bm k}_{2},{\bm \Delta})}{M^2}+\cdots\,,
\end{split}
\ee
where we have made the replacement $\Delta^i\Delta^j \to \frac{{\bm \Delta}^2}{2}\delta^{ij}$ using the symmetry of the ${\bm \Delta}$-integral.\footnote{In principle,  $C_g({\bm k},{\bm \Delta})\propto S({\bm k},{\bm \Delta})$ can depend on $\phi_\Delta$. However, the angular dependent part (the elliptic gluon distribution) is in general small in magnitude \cite{Hagiwara:2016kam} and can be neglected in the present analysis. } 
We now make a crucial use of the formula \cite{Boer:2018vdi,Bhattacharya:2024sno}
\be
\frac{x C_g(x,{\bm k},{\bm \Delta})}{M^2} \approx - \frac{ N_c}{8\pi^4\alpha_s} S(x,{\bm k},{\bm \Delta})\,, \label{cgc}
\ee
valid when $x\ll 1$. 
We thus arrive at 
\be
\begin{split}
 \frac{d\sigma_{\Delta q \Delta q}}{dy_1dy_2 d{\bm P}_{1} d^2 {\bm P}_{2}}& = \frac{1}{2}\frac{1}{(2\pi)^6} \frac{\cos(\phi_{P_1}-\phi_{P_2})}{|{\bm P}_{1}| |{\bm P}_{2}|} \int \frac{d z_1}{z^2_1} D(z_1) \int\frac{d z}{z^2_2} D(z_2) z_1z_2
 \\
& \times \int d^2 {\bm \Delta} {\bm \Delta}^{2} x_1 x_2 F_{\Delta q \Delta q}(x_1,x_2,{\bm \Delta}) S(x_1',{\bm k}_{1},-{\bm \Delta}) S(x_2',{\bm k}_{2},{\bm \Delta})\,,
\end{split}
\label{eq:final}
\ee
where we used $x_{1,2}x'_{1,2}={\bm k}_{1,2}^2/s$ and kept only the cosine term. 

The minus sign in (\ref{cgc}) indicates that the spin and orbital angular momentum of individual gluons are anti-aligned. This has deep theoretical implications connecting with quantum entanglement \cite{Bhattacharya:2024sno,Hatta:2024lbw}. Unfortunately, the sign information  is lost in (\ref{eq:final}) since the distribution is squared. Still, the existence of the strong spin-orbit correlation leads to a new source of `directed flow' proportional to the dipole $S$-matrix squared.   

It is interesting to note that the same cosine correlation  derives from similar manipulations  
 $({\bm k}_1\cdot {\bm \Delta})({\bm k}_2\cdot {\bm \Delta})\to \frac{1}{2}{\bm \Delta}^2 ({\bm k}_1\cdot {\bm k}_2)$ in the odderon case (\ref{odd}) and $({\bm k}_1\times {\bm \Delta})({\bm k}_2\times {\bm \Delta}) 
\to \frac{1}{2}{\bm \Delta}^2 ({\bm k}_1\cdot {\bm k}_2)$ in the spin-orbit case. 
Given that the (hard QCD) odderon has not been observed in experiments, it is expected that the odderon amplitude squared $OO\sim ({\rm Im}S)^2$ in (\ref{odd}) is significantly smaller (cf. \cite{Dumitru:2022ooz}) than the Pomeron amplitude squared $ ({\rm Re}S)^2$ in (\ref{eq:final}).  On the other hand, $F_{\Delta q\Delta q}$ is presumably smaller in magnitude than $F_{qq}$, although we do not know how much smaller. In the next section we provide a numerical estimate of the cosine asymmetry in terms of the directed flow $v_1$ coefficient.

\section{Directed flow $v_1$}

Experimentally, in $pp$ and $pA$ collisions,  the flow coefficients $v_n$ are extracted from the two-particle (two-hadron) correlation 
\beq
\frac{dN}{d\phi_1d\phi_2}\propto 1+2\sum_n v_{n,n}\cos n\Delta \phi.
\eeq
After subtracting `nonflow' contributions such as short range (in rapidity) correlation on the near-side $\Delta\phi\approx 0$ and the recoil  contribution (from $2\to 2$ scatterings) on the away-side $|\Delta\phi|\approx \pi$ \cite{ATLAS:2014qaj,Borghini:2000cm}, $v_{1,1}$  at forward rapidities $y_{1,2}\gg 1$  will be dominated by the spin-orbit correlation    given by the ratio between (\ref{eq:final}) and (\ref{double}) 
\beq
v_{1,1}({\bm P}_{1,2},y_{1,2}) =\frac{1}{4|{\bm P}_1|| {\bm P}_{\rm 2}|} \frac{ \int \frac{dz_1}{z^2_1}D(z_1)\int \frac{dz_2}{z^2_2}D(z_2)\int d^2{\bm \Delta}z_1z_2 x_1x_2 F_{\Delta q\Delta q} (x_1,x_2,{\bm \Delta}) {\bm \Delta}^2 S(x'_1,{\bm k}_1,{\bm \Delta}) S(x'_2,{\bm k}_{\rm 2},{\bm \Delta})  }{ 
\int \frac{dz_1}{z_1^2}D(z_1)\int \frac{dz_2}{z^2_2}D(z_2)  \int d^2{\bm \Delta} x_1x_2F_{qq}(x_1,x_2,{\bm \Delta})   S(x'_1,{\bm k}_1,{\bm \Delta}) S(x'_2,{\bm k}_{\rm 2},{\bm \Delta}) }. \label{v11}
\eeq 
 From this,  $v_1$ of the `trigger' particle is calculated as 
\beq
v_1 ({\bm P},y)= \frac{v_{1,1}({\bm P},{\bm P}_{\rm ref},y,y_{\rm ref})}{ \sqrt{|v_{1,1}({\bm P}_{\rm ref},{\bm P}_{\rm ref},y_{\rm ref},y_{\rm ref})|}} .
\label{v1def}
\eeq
In order for the `flow' interpretation to be viable, $v_1$ should be independent of the reference momentum ${\bm P}_{\rm ref}$ and $y_{\rm ref}$ of the `associate' particle. This is however not guaranteed in general, and will be investigated below. 

We employ a Gaussian model for the double parton distributions 
\beq
F_{qq}(x_1,x_2,{\bm \Delta}) = F_{qq}(x_1,x_2)e^{-{\bm  \Delta}^2/2\sigma^2}, \quad F_{\Delta q \Delta q}(x_1,x_2,{\bm \Delta}) = F_{\Delta q\Delta q}(x_1,x_2)e^{-{\bm \Delta}^2/2\sigma^2},
\eeq
The width  can be different for the two distributions, but for simplicity we adopt a common value $\sigma =0.26$ GeV  \cite{Chang:2012nw}.  In the absence of correlation, $F_{qq}(x_1,x_2)\approx q(x_1)q(x_2)$ up to the kinematical constraint $x_1+x_2<1$, but  $F_{\Delta q\Delta q}(x_1,x_2)$ does not have such a reference and should be entirely modeled. 
As for the dipole $S$-matrix, we assume that the ${\bm \Delta}$-dependence is factorized as 
\be 
S(x',{\bm k},{\bm \Delta}) = S(x',{\bm k}) e^{-R_A^2 {\bm \Delta^2}/4}\,.
\label{Sfact}
\ee
where  $R_A = 1.2 A^{1/3}$ fm, with $A$ being the atomic number, is the target radius.
In this case the ${\bm \Delta}$ integral can be performed and  (\ref{v11}) becomes
\beq
v_{1,1}({\bm P}_{1,2},y_{1,2}) =\frac{1}{4|{\bm P}_1|| {\bm P}_{\rm 2}|} \frac{2}{\frac{1}{\sigma^2}+R_A^2} \frac{ \int \frac{dz_1}{z_1^2}D(z_1)\int \frac{dz_2}{z_2^2}D(z_2) z_1z_2x_1x_2 F_{\Delta q\Delta q} (x_1,x_2) S(x_1',{\bm k}_1)S(x_2',{\bm k}_2)}{ 
\int \frac{dz_1}{z_1^2}D(z_1)\int \frac{dz_2}{z^2_2}D(z_2)  x_1x_2F_{qq}(x_1,x_2) S(x_1',{\bm k}_1)S(x_2',{\bm k}_2)}. 
\label{v11mod}
\eeq
 From \eqref{v11mod}, we expect the result is not very sensitive to the details of the functional form of $S(x',{\bm k})$. Indeed, the only difference in the integrands of the numerator and the denominator is the factor $z_1z_2$. If we were to completely ignore quark fragmentation, the result would be simply
\beq
 v_1 ({\bm P},y)=\frac{1}{\sqrt{2}|{\bm P}|}\frac{{\rm sgn}(F_{\Delta q\Delta q})}{\sqrt{\frac{1}{\sigma^2}+R_A^2}} \sqrt{\frac{|F_{\Delta q\Delta q}(x,x_{\rm ref})|}{F_{qq}(x,x_{\rm ref})}} \overset{A=1}{=} \frac{ 0.098}{|{\bm P}|}{\rm sgn}(F_{\Delta q\Delta q}) \sqrt{\frac{|F_{\Delta q\Delta q}(x,x_{\rm ref})|}{F_{qq}(x,x_{\rm ref})}} ,
\label{est}
\eeq
where $|{\bm P}|$ is in units of GeV on the right hand side. This result is a direct consequence of the relation \eqref{cgc} valid in the small-$x$ limit and leads to  $v_1$ that is independent of $S(x',{\bm k})$. We immediately notice a power-law $1/|{\bm P}|$ falloff. This is distinct from the usual expectation from hydrodynamics $v_1\propto |{\bm P}|$, and also  from the odderon scenario   \cite{Boer:2018vdi} which predicts a sharp peak around $|{\bm P}|\sim 0.2$ GeV.
 Concerning the atomic number dependence, we find  $v_1\sim 1/R_A\sim A^{-1/3}$. This is similar to the finding in  \cite{Boer:2018vdi}. 

Switching on the quark fragmentation, \eqref{est} should remain a reasonable result for large $|{\bm P}|$. To confirm this numerically, we focus only on one representative channel $uu \to \pi^+\pi^+$. At the moment, including other channels $ud$, $dd$, etc., does not necessarily improve the calculation due to large uncertainties in modeling  the double PDFs.  
Even in the $uu$ channel, predictions in the literature vary wildly, with the ratio $r=F_{\Delta u\Delta u}/F_{uu}$ being both negative 
\cite{Chang:2012nw,Rinaldi:2014ddl} and positive \cite{Peng:2024qpw}. The `max scenario' $F_{\Delta q\Delta q}=\pm F_{qq}$, saturating the positivity bound, has also been considered    \cite{Diehl:2014vaa,Echevarria:2015ufa}. In order to assess the maximal impact of our mechanism, we combine the model for $F_{qq}$ from \cite{Gaunt:2009re} with the maximally negative scenario 
\beq
F_{uu}(x_1,x_2) = \frac{(1-x_1-x_2)^2}{(1-x_1)^{2.5}(1-x_2)^{2.5}}u(x_1)u(x_2)\theta(1-x_1-x_2) = -F_{\Delta u \Delta u}(x_1,x_2).
\label{dpdfmod}
\eeq
The prefactor smoothly enforces the kinematical constraint $x_1+x_2<1$ while (approximately) satisfying sum rules. For the dipole $S$-matrix, we use the GBW model \cite{Golec-Biernat:2017lfv} 
and another model from \cite{Lappi:2013zma} obtained by solving the Balitsky-Kovchegov  equation \cite{Balitsky:1995ub,Kovchegov:1999yj}, labeled as the MV$^e$ model. In the numerical  computation we take advantage of the recent fit in \cite{Salazar:2021mpv}.   For the PDF we use the fit from \cite{Dulat:2015mca} and for the FF we use \cite{Ethier:2017zbq}.

\begin{figure}[htb]
  \begin{center}
  \includegraphics[scale = 1]{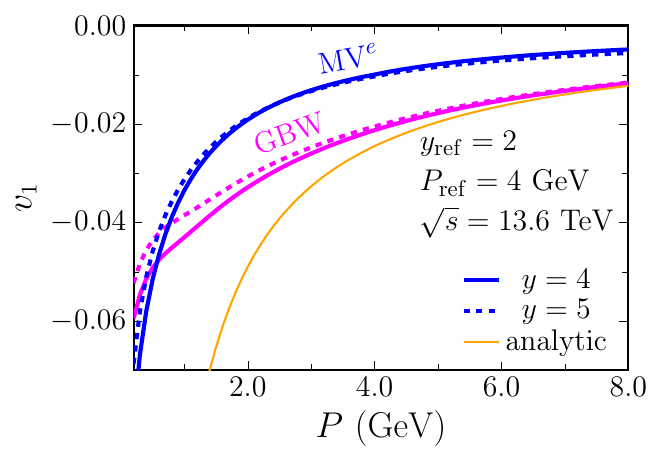}
  \end{center}
  \caption{$v_1$ as a function of $|{\bm P}| = P$ for $pp$ collisions at the LHC with $\sqrt{s} = 13.6$ TeV  with $y_{\rm ref} = 2$ and $|{\bm P}_{\rm ref} |= 4$ GeV. The orange  line is  from \eqref{est}. The magenta lines are obtained using the GBW model, while the blue lines are obtained using the MV$^e$ model. The solid curves are for $y=4$ and the dashed curves are for $y=5$.
  }
  \label{fig1}
\end{figure}
The result in $pp$ collisions at the LHC energy  $\sqrt{s}=13.6$ TeV  is shown in Fig.~\ref{fig1}  with $|{\bm P}_{\rm ref}|=4$ GeV, $y_{\rm ref}=2$ and $y\ge y_{\rm ref}+2=4$. When $|{\bm P}|\sim 1$ GeV, the two models give similar values which are substantially    smaller than \eqref{est}. Increasing $|{\bm P}|$,  the GBW model approaches \eqref{est}, while $|v_1|$ in the MV$^e$ model is about a factor of 2 smaller. The  $y$-dependence is weak up to $y<5$ which is the limit of coverage at the LHCb experiment. We remind the reader that this is the upper limit of $|v_1|$  assuming $r=F_{\Delta u\Delta u}/F_{uu}=-1$.  If $|r|\sim 0.1$, for example, then $v_1$ is scaled down by a factor $\sqrt{0.1}\approx 1/3$. Still, it  remains at the percent level in the low-$|{\bm P}|$ region.

\begin{figure}[htb]
  \begin{center}
  \includegraphics[scale = 0.7]{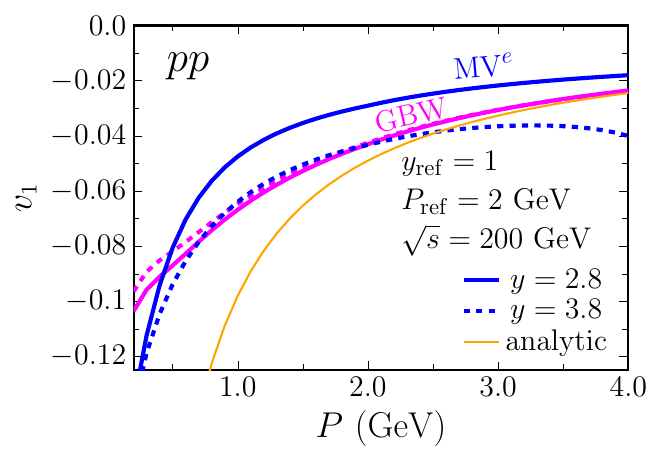}
  \includegraphics[scale = 0.7]{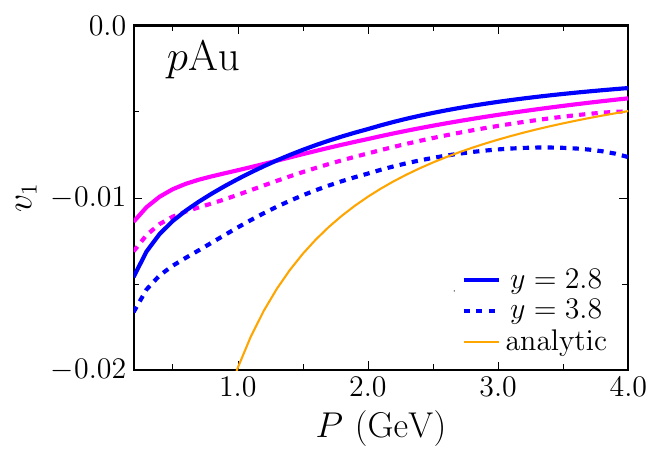}
  \end{center}
  \caption{$v_1$ as a function of $|{\bm P}|=P$ for $pp$ (left) and $p$Au (right) collisions at RHIC with $\sqrt{s} = 200$ GeV with $y_{\rm ref} = 1$ and $|{\bm P}_{\rm ref} |= 2$ GeV. Line styles are the same as in Fig.~\ref{fig1}.}
  \label{fig2}
\end{figure}

Fig.~\ref{fig2} show the results at the RHIC energy $\sqrt{s} = 200$ GeV for $pp$ collisions (left) and proton-gold ($p$Au) collisions with $A=197$ (right). $|v_1|$ is somewhat larger than at the LHC and closer to \eqref{est}. We use $y_{\rm ref}=1$ and $2.8<y<3.8$ following the STAR experiment \cite{STAR:2015kak}. The trends with $y$ and $|{\bm P}|$ are the same as in Fig.~\ref{fig1}, although the magnitude $|v_1|$ is smaller in $pA$ collisions as expected.  Interestingly, we find that $v_1$ at $y=3.8$ in the MV$^e$ model exceeds  (\ref{est}) in magnitude in the high transverse momentum region close to the kinematical limit. This is so even though the ratio of integrals in (\ref{v11mod}) is always less than unity (since $z_{1,2}<1$) and can be understood from the $x_1 + x_2 < 1$ constraint that affects the numerator in \eqref{v1def}.  The same behavior occurs for the LHC kinematics in the highest $|{\bm P}|$ region, and also in the GBW model both at RHIC and the LHC, albeit in a less conspicuous way.  

We have numerically checked that, to an excellent approximation, $v_1$ is independent from $|{\bm P}_{\rm ref}|$ and $y_{\rm ref}$. There is only a small $|{\bm P}_{\rm ref}|$-dependence in the region $|{\bm P}_{\rm ref}| \lesssim 2$ GeV. The dominant ${\bm P}_{\rm ref}$ dependence in \eqref{v11} automatically factorizes through the overall term $\sim 1/|{\bm P}||{\bm P}_{\rm ref}|$. The remaining $|{\bm P}_{\rm ref}|$-dependence is very weak because it comes from the ratio of the integrals \eqref{v11} and also because we factorized the ${\bm k}$ and ${\bm \Delta}$ dependencies through \eqref{Sfact}. Likewise, $v_1$ is approximately independent of $y_{\rm ref}$.

Turning to the experimental data, in proton-lead ($p$Pb) collisions at the LHC, around midrapidity $|y|<2.5$ and $2<|y-y_{\rm ref}|<5$, the ATLAS collaboration found   $v_1\approx a|{\bm P}|$ with $a<0$ at small-$|{\bm P}|$  
\cite{ATLAS:2014qaj}. $v_1$ then flips signs around $|{\bm P}|\sim 1.5$ GeV and continues to grow towards the  high-$|{\bm P}|$ region. Such a behavior is seemingly consistent with the  hydrodynamic evolution of the final state  \cite{Teaney:2010vd}. $v_1$ was also measured by the STAR collaboration in deuteron-gold ($dA$) collisions (without recoil subtraction)  
\cite{STAR:2015kak} selecting trigger  particles  to be in the midrapidity region $|y|<1$.

Our calculation requires both the trigger and associate particles  to be in the forward rapidity region $y$, $y_{\rm ref}>0$ (ideally $y$, $y_{\rm ref}\gg 1$) while keeping $|y-y_{\rm ref}|\gtrsim 2$. The LHCb experiment, with a  wide acceptance at forward rapidity  $2<|y|<5$, is suited to  this purpose. The two-particle angular correlation has been  already measured \cite{LHCb:2015coe,LHCb:2023png}. It would be interesting to reanalyze the data with a focus on $v_1$ and its $|{\bm P}|$ dependence.

\section{Conclusions}

In this paper, we have found a novel mechanism to generate $v_1$, the directed flow coefficient, in the forward rapidity region of $pp$ and $pA$ collisions. This is a direct consequence of the nonvanishing spin-orbit correlation  for gluons. At small-$x$, the correlation is `maximally' strong $|C_g(x)|\approx |g(x)|$ such that, at the partonic level, $v_1$ does not depend on the details of the dipole $S$-matrix (\ref{est}). While in practice  we have found some model dependence after convoluting with the fragmentation function, the observable offers a unique source of information to constrain the proton's double helicity PDFs $F_{\Delta q\Delta q}$.  In particular, the sign of $F_{\Delta q\Delta q}$ can be inferred from that of $v_1$.

The present work can be extended in  several directions. One can include the scattering of two gluons  from the double gluon helicity PDF of the projectile proton $\langle  (F^{+\mu}\tilde{F}^+_{\mu})(F^{+\nu}\tilde{F}^+_{\nu})\rangle$ (cf. \cite{Boussarie:2018zwg}).  Although this is subleading in the very forward region, it may give an important contribution at realistic rapidities. An interesting observation is that there will be no odderon contribution in this case because the dipole $S$-matrix in the adjoint representation is purely real. The effect of the quark spin-orbit correlation $C_q(x)$ can also be studied considering the process $(q\bar{q}^*\to g)^2$. 
Finally, it is  interesting to study $v_1$ in the forward region of  ultraperipheral collisions (UPCs) or in the photoproduction limit of DIS following 
\cite{Shi:2020djm}. The almost real photon from the projectile  may be treated as a `hadron' with its own double helicity PDFs. This may be explored at the LHC, RHIC and the future Electron-Ion Collider (EIC). 

\section*{Acknowledgments}

We thank Feng Yuan for discussions. S. B. is supported by the Croatian Science Foundation (HRZZ) no. 5332 (UIP-2019-04).
Y.~H. is supported by the U.S. Department
of Energy under Contract No. DE-SC0012704, by LDRD funds from Brookhaven Science Associates, and also by the framework of the Saturated Glue (SURGE) Topical Theory Collaboration.

\bibliography{ref}
\end{document}